\documentclass[article,3p,times,twocolumn, nopreprintline]{elsarticle}
\usepackage{epstopdf}
\usepackage{graphicx}
\usepackage{amssymb}
\usepackage{tabularx}
\usepackage{subcaption} 
\usepackage{caption}

\usepackage{amsmath}
\usepackage{pifont}
\usepackage{supertabular}
\usepackage{eurosym}
\usepackage{xcolor} % to use \textcolor{red}{blabla}
\usepackage{mathtools}
\usepackage{multirow}
\usepackage{adjustbox}
\usepackage{lipsum}% http://ctan.org/pkg/lipsum
\usepackage{multicol}% http://ctan.org/pkg/multicols

\usepackage{appendix}
\usepackage{booktabs}
\usepackage{array}

\usepackage{amsthm}

\usepackage[ruled]{algorithm2e}
\usepackage{algpseudocode}
\usepackage{float}
\usepackage{centernot}
\usepackage[draft=True]{hyperref} % put False when you've finished all your writing
\hypersetup{
     colorlinks=true,
     linkcolor=blue,
     filecolor=blue,
     citecolor=black,      
     urlcolor=cyan,
     }

\begin{document}

\begin{frontmatter}

\title{Ammonia, Methane, Hydrogen and Methanol Produced in Remote Renewable Energy Hubs: a Comparative Quantitative Analysis}
\author[ULIEGE]{Antoine Larbanois$^*$}
\author[ULIEGE]{Victor Dachet$^*$}
\author[ULIEGE]{Antoine Dubois}
\author[ULIEGE]{Raphaël Fonteneau}
\author[ULIEGE,PARIS]{Damien Ernst}
\address[ULIEGE]{Department of Computer Science and Electrical Engineering, Liege University, Liege, Belgium}
\address[PARIS]{LTCI, Telecom Paris, Institut Polytechnique de Paris, Paris, France}
\date{November 2023}

\begin{abstract}

Remote renewable energy hubs (RREHs) for synthetic fuel production are engineering systems harvesting renewable energy where it is particularly abundant. They produce transportable synthetic fuels for export to distant load centers. This article aims to evaluate the production costs of different energy carriers, and includes a discussion on advantages and disadvantages in terms of technical performance. To do so, we extend the study of \citet{berger2021remote} which  focuses on methane (CH$_4$) as energy carrier and introduce three new carriers: ammonia (NH$_3$), hydrogen (H$_2$) and methanol (CH$_3$OH). The four different RREHs are located in the Algerian Sahara desert and must serve  to the load center, Belgium, a constant electro-fuel demand of 10 TWh per year. The modelling and optimisation of these systems are performed using the modelling language GBOML (Graph-Based Optimisation Modelling Language). Our findings reveal that the three new RREHs, each with its respective carrier (ammonia, hydrogen, and methanol), are all more cost-effective than the methane-based system. Ammonia demonstrates the most favourable cost-to-energy exported ratio.

\end{abstract}

\end{frontmatter}

\def\thefootnote{*}\footnotetext{These authors contributed equally to this work.}\def\thefootnote{\arabic{footnote}}

\section{Introduction}

To decarbonise its energy infrastructure, Europe must harness substantial quantities of renewable energy (RE) sources.
Nevertheless, several European nations are struggling with constraints related to the exploitation of locally available RE resources, posing a challenge in meeting the continent's energy demands. 
These limitations are stemming from a confluence of factors, including spatial constraints for RE infrastructure deployment and low-quality RE resources, especially in countries like Belgium which are densely populated \cite{elia2021NetZero}. 

The Remote Renewable Energy Hub (RREH) concept has emerged as a compelling solution in response to these constraints. 
Situated in regions far away from major population centres, RREHs harness the benefits of abundant and cost-effective RE sources, thereby mitigating the shortfall in local renewable resources within Europe. 
These energy production systems convert substantial amounts of electricity from remote renewable resources into high-energy-density molecules, which can be transported to urban load centres. 
These molecules can for example serve as crucial raw materials for industrial sectors or function as dispatchable sources of electricity generation. 

Although there are many open research questions related to these hubs \cite{berger2021remote, fonder2023synthetic, dachet2023co2, verleysen2023where}, a predominant question concerns the types of energy-rich molecules that should preferably be synthesized in those. As a step to answer this question, we conduct a comparison between the methane-based system introduced by \citet{berger2021remote} and three new energy carriers: ammonia (NH$_3$), hydrogen (H$_2$) and methanol (CH$_3$OH), with the aim of identifying a more cost-effective synthetic fuel supply chain.
According to \citet{berger2021remote}, the methane's results showed a cost of approximately 150€/MWh (HHV) with costs estimated at 2030 for a system delivering 10 TWh of e-gas annually.

Our findings reveal that the three new RREHs, each with its respective carrier (ammonia, hydrogen, and methanol), are all more cost-effective than the methane-based system. Ammonia demonstrates the most favourable cost-to-energy exported ratio at 107€/MWh (HHV). This arises from its superior efficiency throughout the studied supply chain. The findings demonstrate that it is less expensive to combine hydrogen with nitrogen taken from the air using an air separation unit system, rather than using carbon dioxide captured by a direct air capture device.

This article is organized as follows: Section 2 introduces the case studies and the optimization framework, Section 3 details the results. 
Lastly, Section 4 serves as the conclusion.

\section{Case studies} \label{sec:case_study}

In this section, we give a detailed overview of the energy systems that are compared in this case study.
We start by detailing the characteristics of the original RREH designed in \cite{berger2021remote}, whose exported energy carrier is methane.
Then, using this case as a reference, we detail the three other scenarios.
For better transparency and comparability, we follow by integrating these four scenarios in the framework proposed by \citet{Dachet2023Taxonomy}.
Finally, we give a brief overview of the optimisation framework that was used to derive the results presented in Section 3.

\subsection{RREH in Algeria}

\begin{figure} [!ht]
    \centering
    \includegraphics[width=0.5\textwidth]{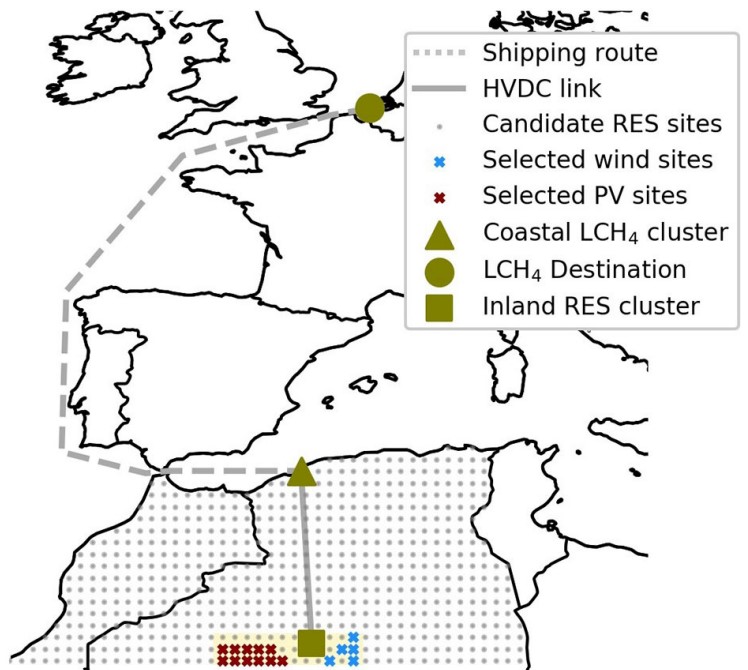}
    \caption{The configuration of the RREH, depicted in this picture sourced from \cite{berger2021remote},} consists of three regions: firstly, the Algerian inland where renewable electricity is produced, secondly, the Algerian coast where the electricity is converted into e-fuel, and finally, Belgium where the e-fuel is delivered.
    \label{fig:RREH ALGERIA}
\end{figure}

In this article, the remote hub system configuration is formed by three distinct geographical areas as show in \autoref{fig:RREH ALGERIA}. 
The Algerian inland is used for harnessing the abundant renewable resource reservoirs through the utilization of photovoltaic panels and wind turbines.
The electricity generated is then channelled with an HVDC interconnection to the coastal region of Algeria, to be converted through different technological processes into e-fuels. 
In each RREH studied, storage units are used to provide flexibility between processes. 
In particular through the use of lithium-ion batteries, gaseous hydrogen tanks as well as tanks of different commodities.
Subsequently, the resultant e-fuels are shipped to the load centre. 
During maritime transport, each energy carrier is considered both as fuel and as a cargo.
The load center's demand remains consistent across all the studied systems: consistently consuming 10 TWh (HHV) of e-fuel each year.

\subsection{Reference scenario: RREH for methane production} \label{Reference scenario}

In this RREH, methane is synthesized by combining hydrogen with carbon dioxide harvested using Direct Air Capture technologies (DAC).

These DAC units operate according to the process proposed by \citet{KEITH20181573}.
This process consists of two interconnected chemical loops. 
In the first loop, $CO_2$ from the atmosphere is captured in contactors using an aqueous solution to form dissolved compounds. 
In the second loop, these compounds then react with $Ca_{2+}$ to produce $CaCO_3$. 
The latter is calcined to release $CO_2$ and $CaO$. 
$CaO$ is then hydrated to produce $Ca(OH)_2$, which regenerates $Ca_{2+}$ by dissolving $Ca(OH)_{2}$. 
Calcination is carried out by burning hydrogen (originally, the fuel is methane) produced by the electrolyzers. 
Moreover, an electricity consumption is needed to powers fans that drive air through the contactors, pumps maintaining the flow of the aqueous solution, and compressors compressing the $CO_2$ flow to 20 bars. 
Additionally, the unit consumes freshwater to create the aqueous solutions, counteract natural evaporation in the contactors, and produce steam for the hydration of $CaO$.

Hydrogen is produced using proton exchange membrane (PEM) electrolyzers \cite{CARMO20134901} (which also produce pure oxygen). 
These units can be supplied with intermittent electricity without any power ramp-up constraints. 
Furthermore, they are supplied with desalinated seawater, coming from reverse osmosis units to produce freshwater \cite{CALDERA2016207}. 
This technology uses a porous membrane to filter seawater, creating a pressure difference that allows the recovery of freshwater on the other side of the membrane. 
Electricity is needed to pump seawater and drive the fresh water. 

The combination of hydrogen and carbon dioxide takes place in a methanation unit following an exothermic reaction, generating both methane and water vapor. 
Furthermore, it is considered that the production of synthetic methane can be flexible, without any power ramp-up constraints. 
However a minimum operating capacity of 40\% of its nominal capacity is assumed \cite{gotz2016renewable}. 
Methane production directly feeds into a liquefaction unit, which increases the volumetric density of the e-fuel for maritime transportation.
In addition to a need for methane, this unit is powered by electricity.
The technology uses compressors and pumps to gradually compress and cool the methane flow, which is then expanded and transformed into liquid through the Joule-Thomson effect.
Power ramp-up constraints are taken into account to ensure uninterrupted operation.
Once delivered to Belgium, the methane undergoes regasification where the heat required for the phase change comes from the combustion of a portion of the methane (2\%) \cite{Dongsha2017}.

\begin{figure*}[p]
    \vspace{-3cm}
    \hspace*{\fill} 
    \includegraphics[angle=90, width=0.45\textwidth]{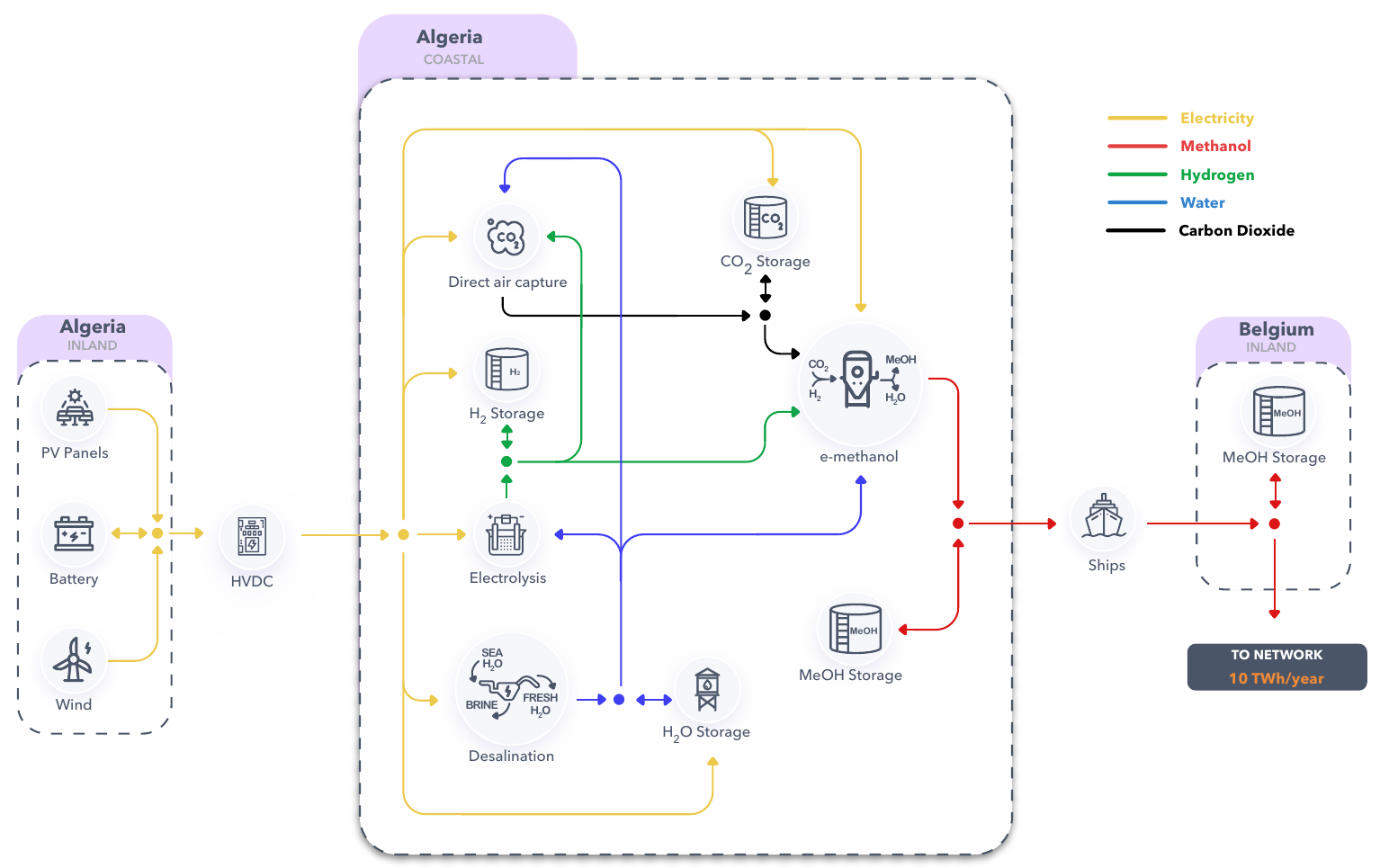}
    \hspace*{\fill} 
    \includegraphics[angle=90, width=0.45\textwidth]{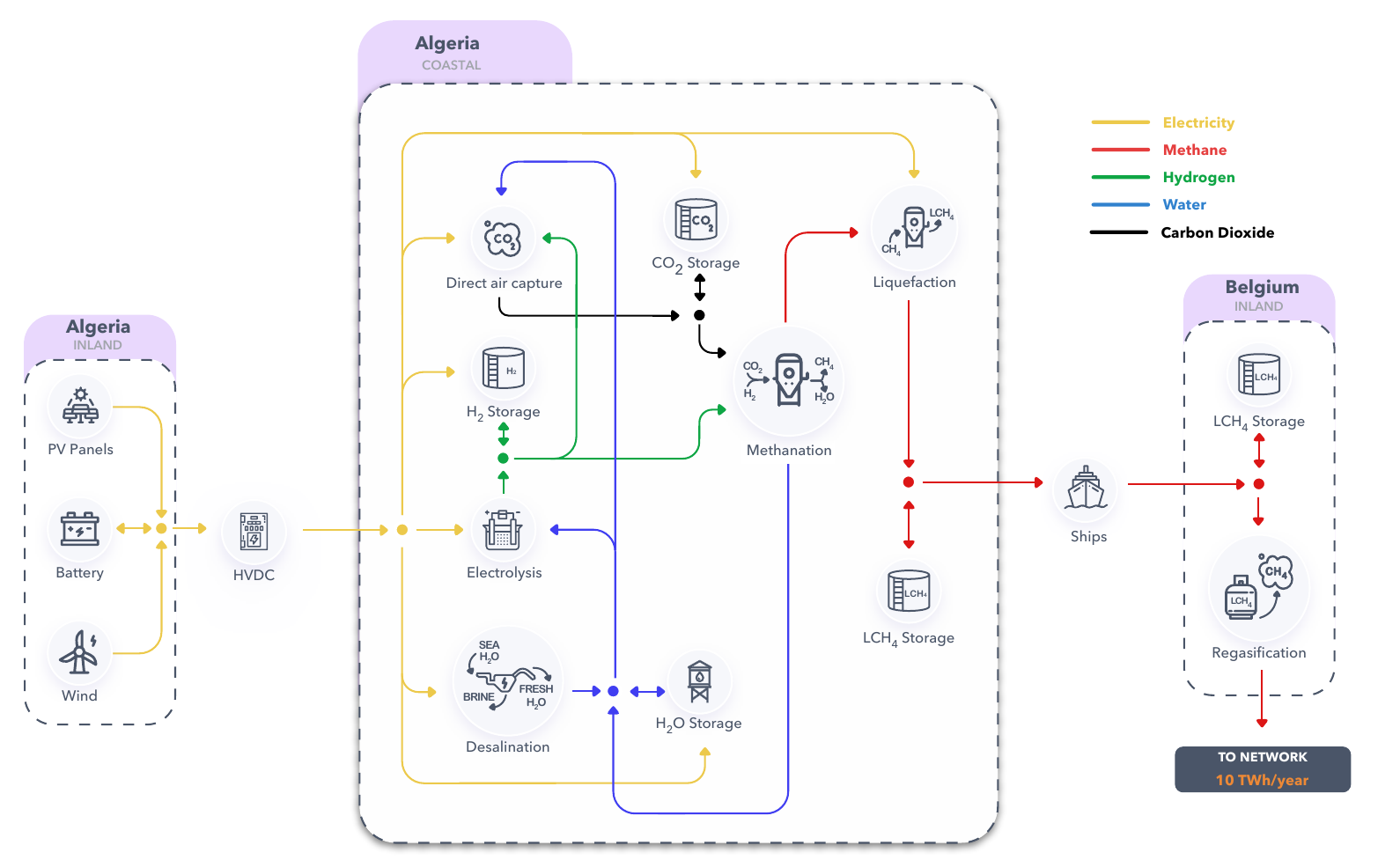}
    \vspace*{\fill} 
    \hspace*{\fill} 
    \includegraphics[angle=90, width=0.45\textwidth]{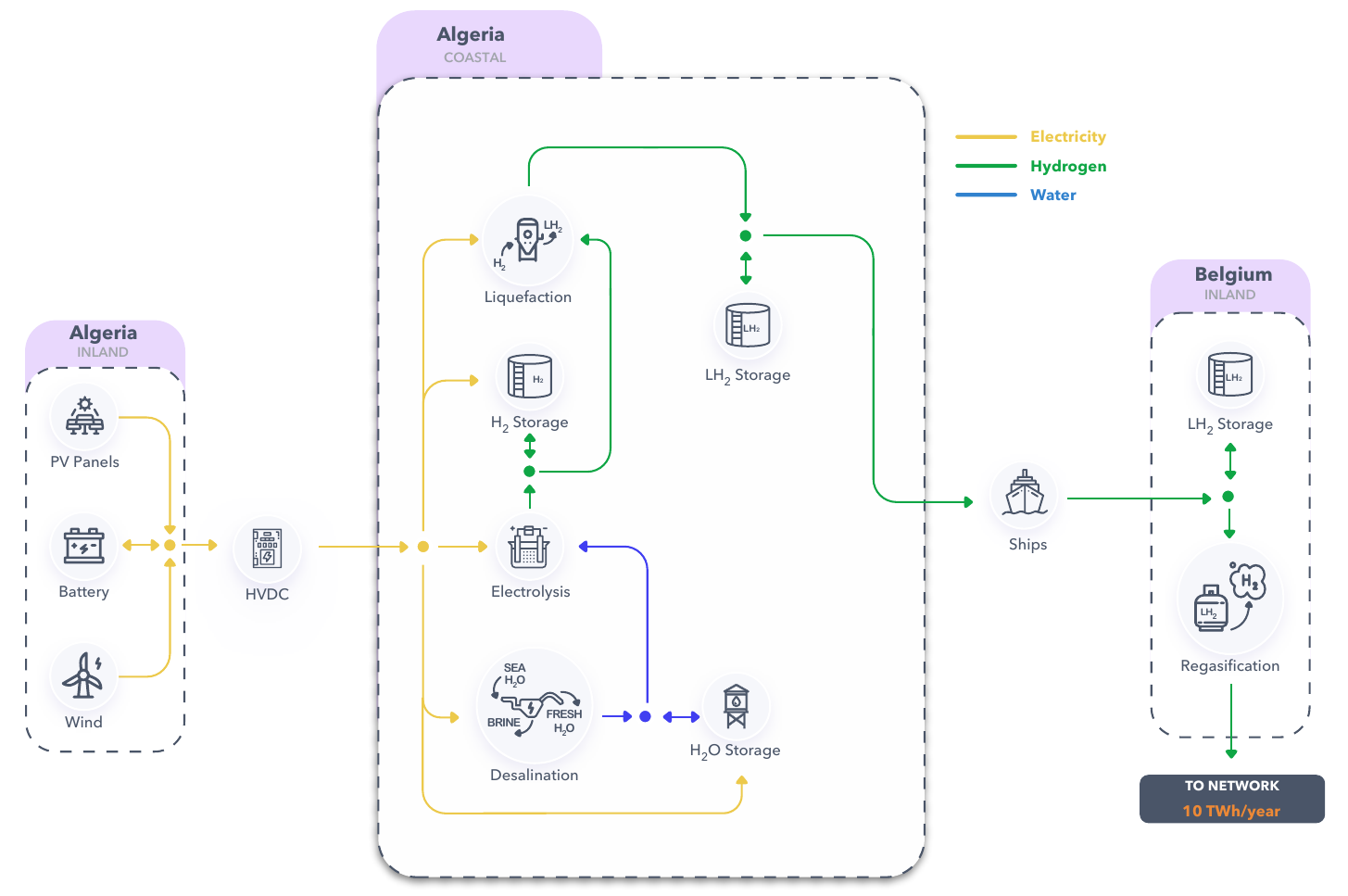}
    \hspace*{\fill} 
    \includegraphics[angle=90, width=0.45\textwidth]{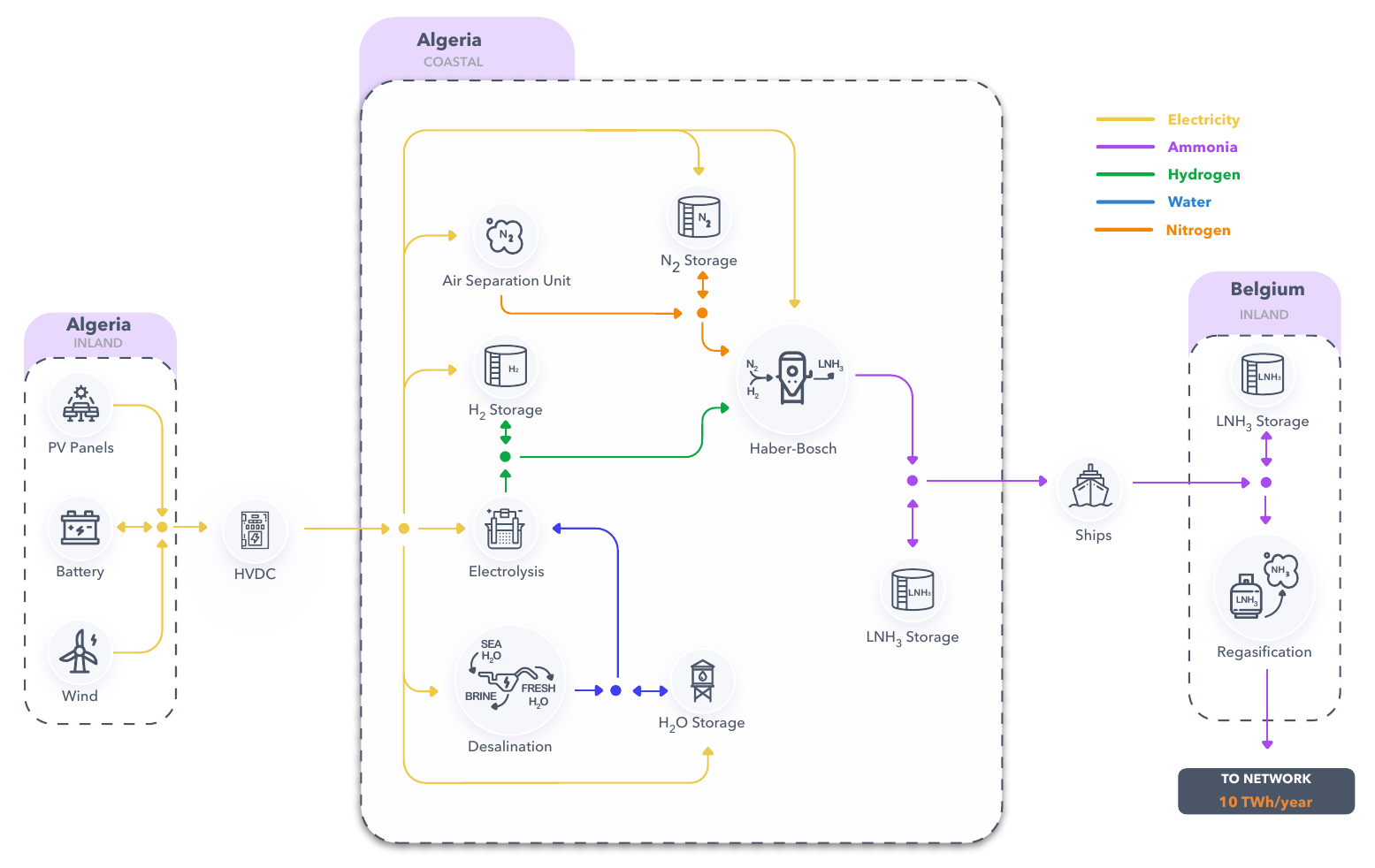}
    \caption{Graph illustrating the 4 RREHs for the 4 potential energy carrier candidates. Each technology is  depicted as a node, and each connection is denoted by a hyper-edge.}
    \label{fig:All_RREHs}
   
\end{figure*}

All the mentioned technologies are interconnected, and these connections are symbolized by various commodities, including electricity, hydrogen, water, carbon dioxide, and methane.
\autoref{fig:All_RREHs} depicts a graphical representation of the interconnections and the technologies involved in the RREH's methane production process.

\subsection{RREH for ammonia, hydrogen and methanol production}

In the ammonia RREH, the hydrogen produced by electrolyzers is associated with pure nitrogen. 
For this purpose, an air separation unit is used to separate ambient air into nitrogen, oxygen, and argon through cryogenic distillation \cite{dea2023renewable}. 
The unit relies on electricity to operate its pumps and compressors.
Furthermore, power ramp-up constraints are used to force the unit to operate continuously. 
The Haber-Bosch process is used to synthesize ammonia (in liquid form) by combining nitrogen and hydrogen under high pressure and high temperature using a catalyst. 
The existence of a "Hot standby" mode which allows maintaining the reactor at an optimal temperature when its supply is insufficient induces a minimum operating threshold of 20\% \cite{dea2023renewable}, without power ramp-up constraints.
During delivery in Belgium, ammonia also undergoes regasification with an energy loss of approximately 2\% of the energy content (same assumption as methane).

In the hydrogen RREH, hydrogen is not associated with any other molecule. 
During its production via electrolyzers, hydrogen is directly liquefied and then transported by sea. 
The liquefaction unit operates continuously and the amount of electrical energy required to liquefy 1 kg of hydrogen is estimated at 12 kWh \cite{dnv2020study}, resulting in an energy loss of approximately 30\%. 
Hydrogen undergoes regasification in Belgium, following the same process as previous e-fuel. 
However, no energy loss is assumed during this process \cite{dnv2020study}.

Finally, the methanol model is quite similar to the methane model (described in Subsection 2.2), with the difference that it does not require liquefaction or regasification units since methanol is liquid at the ambient temperature. 
The process of converting carbon dioxide and hydrogen into methanol is an exothermic reaction that occurs at a temperature of approximately 250°C. 
Additionally, methanol synthesis requires steam at 10 bars and heated to 184°C \cite{dea2023renewable}. 
This steam is modeled as a water consumption (water production at the reactor outlet deducted) and a quantity of burned hydrogen. 
In addition, electricity is necessary to operate auxiliary equipment
Furthermore, it is estimated that the synthesis unit has also a "Hot standby" mode that allows maintaining the reactor at the right temperature and pressure conditions. 
This sets a minimum operating constraint at 10\% \cite{dea2023renewable} and allows quick power variations with constraints.

In Figure \ref{fig:All_RREHs}, you can find representations of all technologies and the connections that link them across the various RREHs.

\subsection{Inclusion in the RREH taxonomy}

To improve the understanding of the key discrepancies between hubs, an RREH can be categorized employing the taxonomy detailed in \citet{Dachet2023Taxonomy}.
This taxonomy is instantiated across the four hubs developed in this study. 
It allows to characterize an RREH with:

\begin{itemize}
    \item $\mathcal{L}$: A set of locations associated  with the technologies in the hub. 
    \item $\mathcal{G}$: A graph formed by technologies and commodity flows. These technologies are situated in locations depicted in $\mathcal{L}$.  
    \item $\mathcal{C}$: Denotes the set of exchanged and produced commodities within the graph.
    \item $\mathcal{I}$: The set of imported commodities in the hub. 
    \item $\mathcal{E}$: The set of exported commodities to the load center.
    \item $\mathcal{B}$: The set of byproducts commodities. These byproducts are commodities output by a technology that are not fully exploited by another technology - they could be however partially exploited.
    \item $\mathcal{O}$: The set of locally exploited opportunities. These locally exploited opportunities are commodities that supply a local demand (different than the load center).

\end{itemize}

\begin{table*}[ht!]
\centering
\begin{tabular}{p{1.5cm}|p{3cm}|p{3cm}|p{3cm}|p{3cm}}
 RREH & Methane & Methanol & Hydrogen & Ammonia\\
\hline 
$\mathcal{L}$ &  \multicolumn{4}{p{12cm}}{\{Sahara desert, Algerian coast\}} \\
\hline
$\mathcal{G}$ &  \multicolumn{4}{p{12cm}}{All Technologies and commodity flows are represented in \autoref{fig:All_RREHs}}\\
\hline
$\mathcal{C}$ &  \{Electricity, H$_2$, H$_2$O, CO$_2$, CH$_4$, O$_2$, heat\} & \{Electricity, H$_2$, H$_2$O, CO$_2$, CH$_3$OH, O$_2$, heat\} & \{Electricity, H$_2$, H$_2$O, O$_2$, heat\} & \{Electricity, H$_2$, H$_2$O, N$_2$, NH$_3$, O$_2$, heat, Ar\}\\
\hline
$\mathcal{I}$ &  \multicolumn{4}{p{12cm}}{\{air, sea water\}}\\ 
\hline
$\mathcal{E}$ &  \{CH$_4$\}   &  \{CH$_3$OH\}   &  \{H$_2$\}  & \{NH$_3$\}\\
\hline
$\mathcal{B}$ & \multicolumn{3}{|p{9cm}|}{\{heat, O$_2$\}} & \{heat, O$_2$, Ar \}\\ 
\hline
$\mathcal{O}$ & \multicolumn{4}{p{12cm}}{ $\emptyset$}\\ 
% $\mathcal{O}$ & \{$H_2O$, electricity, H$_2$, CH$_4$, $CO_2$ \} & \{$H_2O$, electricity, H$_2$, $CO_2$, CH$_3$OH \}& \{$H_2O$, electricity, H$_2$\}& \{$H_2O$, electricity, H$_2$, NH$_3$, $N_2$\} 

\end{tabular}
\caption{Table characterizing the different RREHs according to the taxonomy presented by \citet{Dachet2023Taxonomy}}
\label{tab:taxo}
\end{table*}

In order to enhance the readability of the sets describing the hubs, some notational liberties have been taken compared to the taxonomy of \citet{Dachet2023Taxonomy}. The sets are all listed in \autoref{tab:taxo}. 

As depicted in \autoref{tab:taxo}, the sets characterizing the hubs share significant similarities, even if they relate to different export commodities. Nevertheless, technologies such as DAC and ASU introduce new flows that generate different exchanged commodities than the exported product, such as nitrogen or carbon dioxide. Moreover, it is worth noticing that ASU also enables the production of an additional byproduct, namely argon.

In addition, this taxonomy highlights distinct commodities across the different RREHs. An example of common commodity across all the hubs is the water produced by the desalination units. One could integrate a demand of water close to the hub. This production of drinking water would extend the set of exploited local opportunities. Besides, it may help to face water scarcity in the phase of changing climate patterns and population growth.

\subsection{Optimising the system with GBOML}

Following the optimisation framework introduced by \citet{berger2021remote}, the studied energy systems are represented in a hyper-graph format.
Within this format, interconnected nodes refer to subsystems, such as different technologies, facilities, or processes. 

Each node represents an optimization sub-problem, where optimisation procedures are executed to determine the optimal values of capital expenditure (capex) and operating expenditure (opex) parameters for the various technologies. 
This optimisation process is conducted while meeting the energy demand from the load center.

Nodes are defined by internal variables (such as capacities), external variables (such as commodity production), parameters (such as capex/opex), by constraints and objective functions. 
The connections between these nodes are established through hyper-edges, which represent flows of commodities. They impose conservation constraints on each commodity.  

Furthermore, the models are constructed using the Graph-Based Optimization Modeling Language (GBOML), a language specifically designed for modeling the optimization of multi-energy systems based on graphs, as introduced in \citet{miftari2022gboml}.

The global parameters used for modelling were retained from the methane-based model to maintain a comparable foundation with the new carriers.
These parameters have a capital cost rate of 7\% and a temporal horizon of 5-years, at hourly resolution. 
Moreover, different economic and technical parameters (2030 estimate) used for node modelling are gathered in \autoref{tab:conversion_nodes_tab_tech} to \autoref{tab:Storage_nodes_stock_tab_eco_stock}, see appendices of this document.

\section{Results and discussions} \label{sec:results}

\begin{figure*}[ht!]
    \hspace*{\fill} 
    \includegraphics[width=0.49\textwidth]{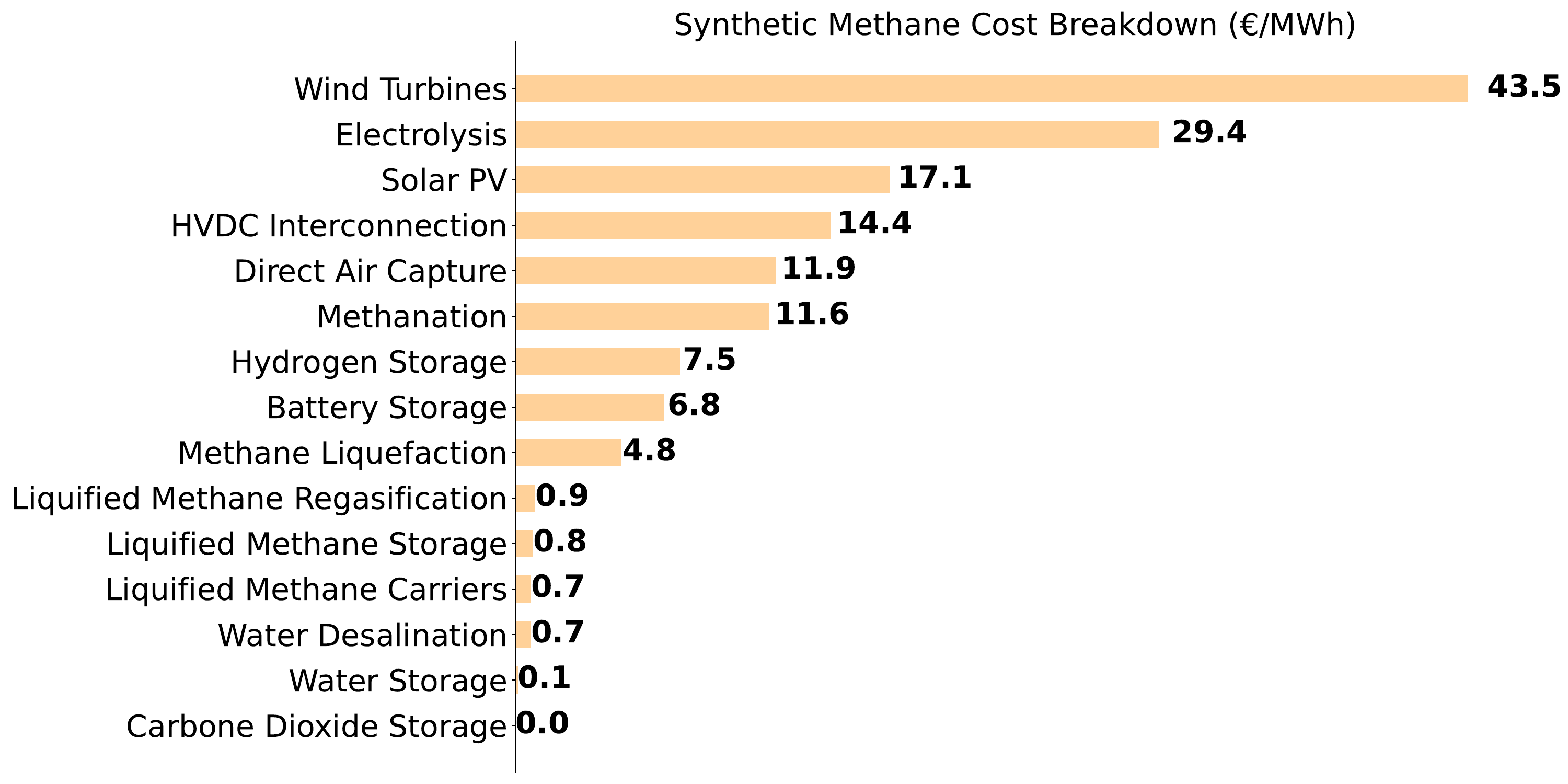}
    %\caption{Contributions to the cost of methane synthesized in an RREH. All contributions roughly sum to 150.2 €/MWh (HHV).}
    %\label{fig:COST_CH4}
    \hspace*{\fill} 
    \includegraphics[width=0.49\textwidth]{{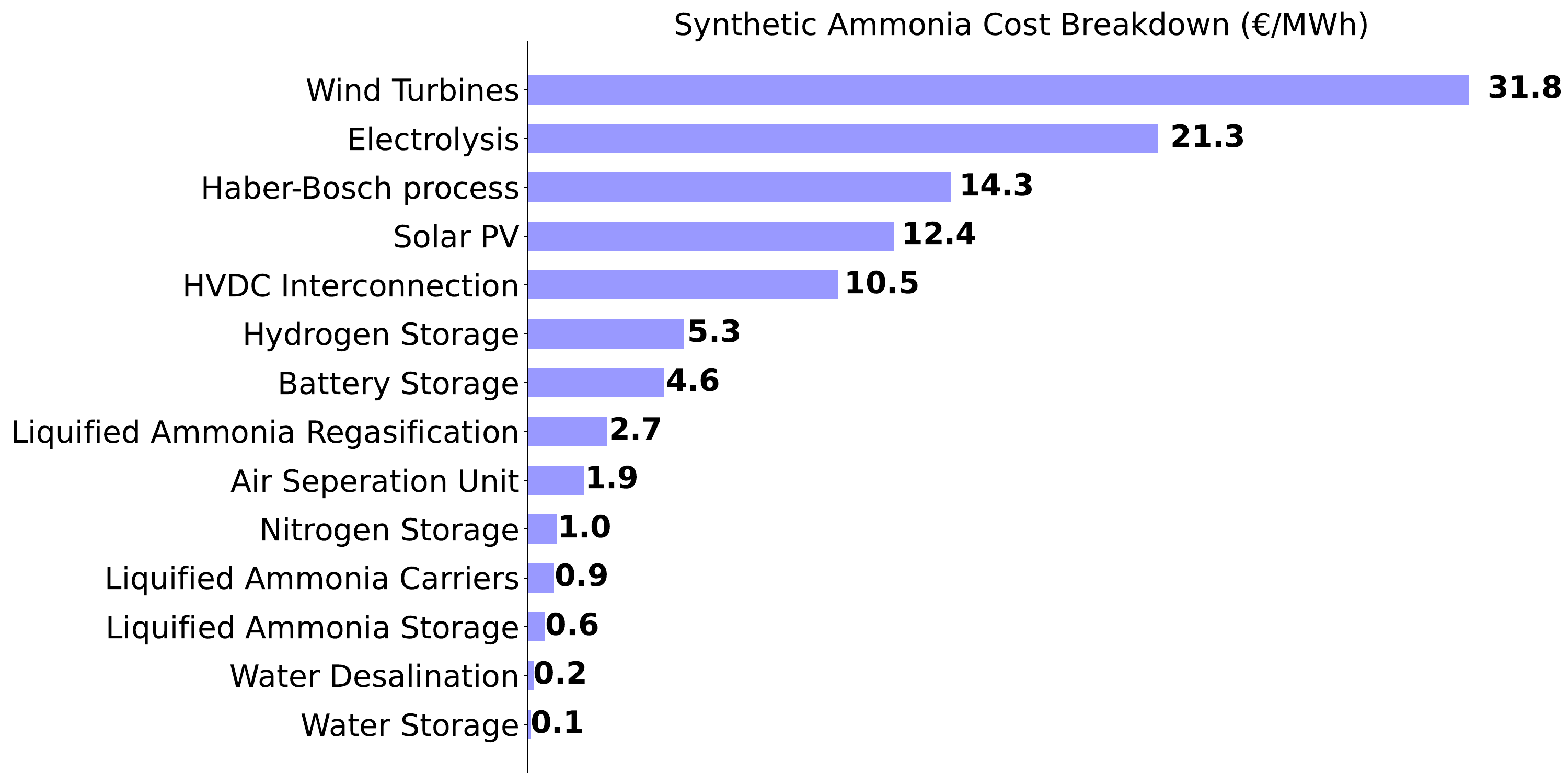}}
    %\caption{Contributions to the cost of ammonia synthesized in an RREH.All contributions roughly sum to 107.4 €/MWh (HHV).}
    %\label{fig:COST_NH3}
    \vspace*{\fill} 
    \hspace*{\fill} 
    \includegraphics[width=0.49\textwidth]{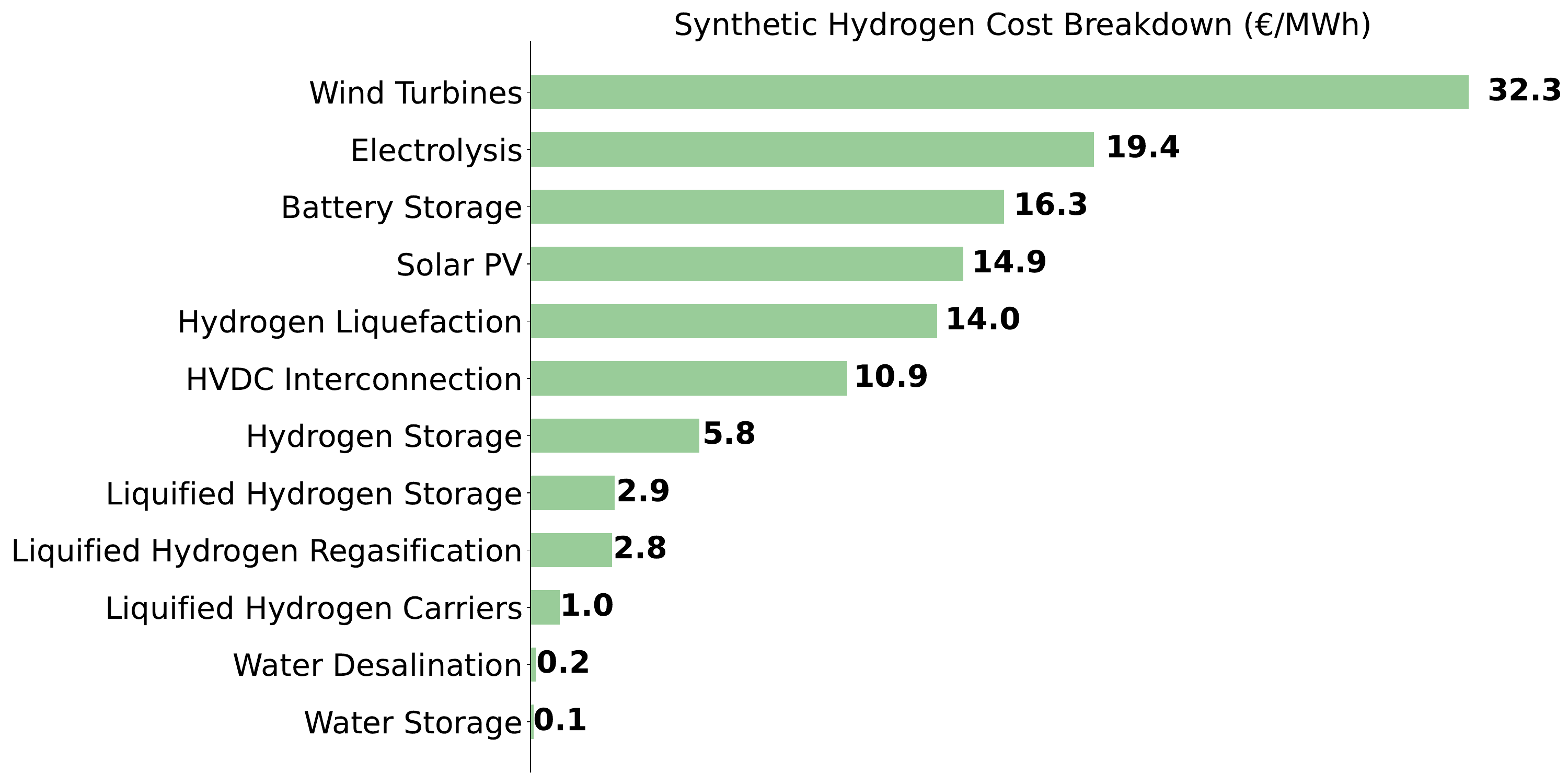}
    %\caption{Contributions to the cost of hydrogen synthesized in an RREH, All contributions roughly sum to 120.5 €/MWh (HHV).}
    %\label{fig:COST_H2}
    \hspace*{\fill} 
    \includegraphics[width=0.49\textwidth]{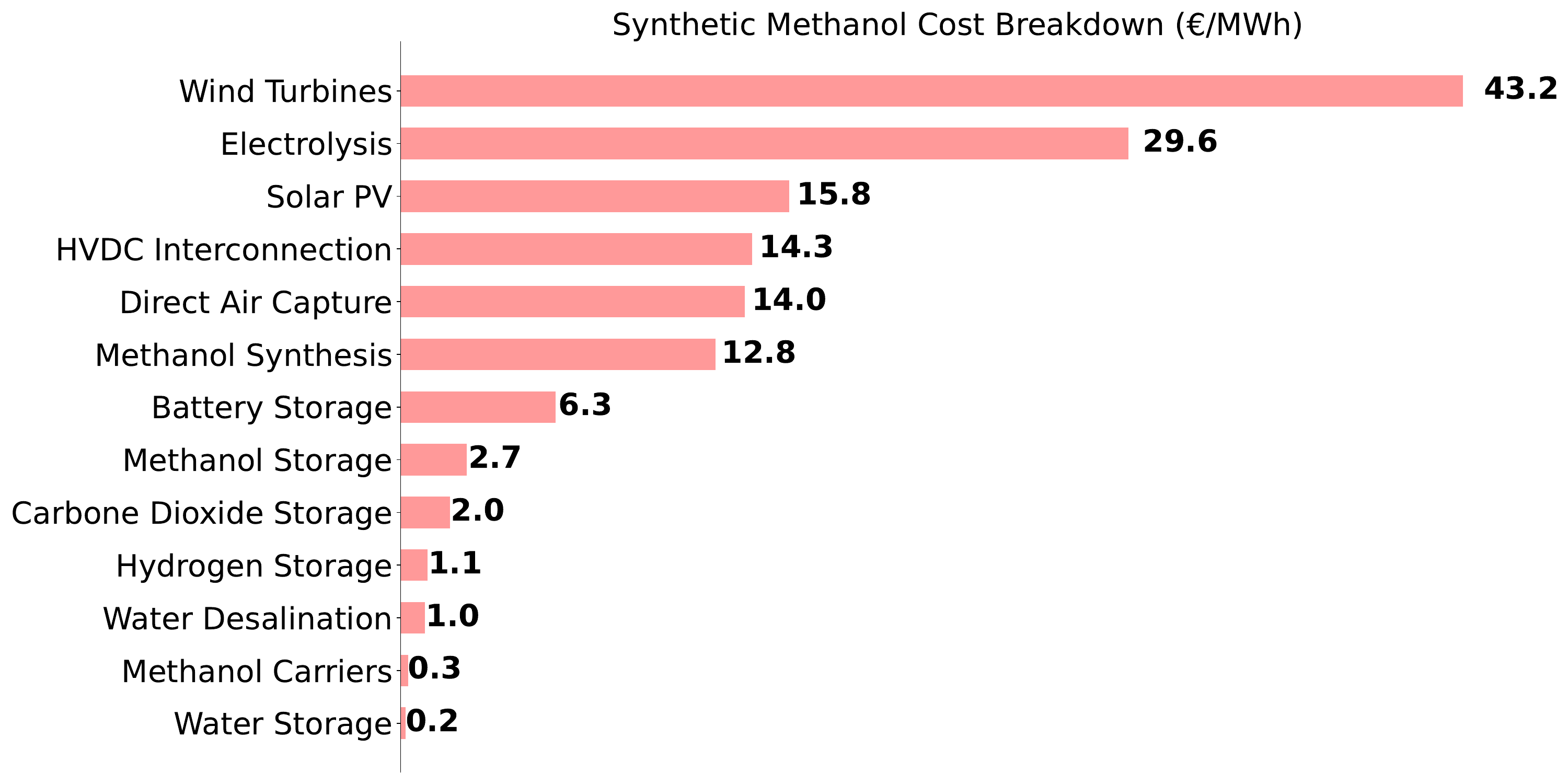}
    \caption{Contributions to the cost of each e-fuel synthesized in an RREH. All contributions roughly sum to 150€/MWh for methane, 107€/MWh for ammonia, 121€/MWh for hydrogen and 143€/MWh for methanol.}
    \label{fig:COST_EFUEL}
\end{figure*}

In this section, we present the costs associated with each vector and we perform a comparative analysis in terms of costs, installed capacities, and overall efficiencies of the vectors. Then, we discuss the results obltained.

\subsection{Results and comparative analysis of e-fuels}

The findings indicate that producing 10 TWh annually of synthetic methane from renewable sources in the Algerian desert results in an overall cost of 7.5 billion euros, equivalent to a rate of 150 €/MWh. 
Shifting the production focus from methane to ammonia brings the total cost down to 5.4 billion euros, corresponding to 107 €/MWh. 
For the cases of hydrogen and methanol, production costs are 6 billion euros and 7.2 billion euros, respectively, translating to 120 €/MWh and 143 €/MWh.

\autoref{fig:COST_EFUEL} depict how technologies contribute to the total production cost of each e-fuel. 
Across all models, wind turbines stand out as the most significant cost component, accounting for approximately 30\% of the total costs, closely followed by electrolyzers with a contribution of around 20\%. 
Overall, technologies related to electricity production, transmission, and storage represent the highest costs, ranging between 55\% and 60\%.

\begin{table}[!ht]
    \centering
    \begin{tabular}{c|c|c|c|c}
         & NH$_3$ & CH$_3$OH & H$_2$ & CH$_4$ \\
        \hline
        Liquid & 103 & 143 & 118 & 146 \\
        Gazeous & 107 & - & 121 & 150 \\
    \end{tabular}
    \caption{Costs in euro per MWh per e-fuels and phase produced in RREHs.}
    \label{tab:costs_comp_sc_0}
\end{table}

The results in \autoref{tab:costs_comp_sc_0} illustrate the costs of each fuel per phase. 
Notably, each gaseous molecule has a lower cost than a gaseous phase due to the efficiency of the regasification unit and its cost to implement.
These findings highlight that among the synthetic fuels considered, ammonia is the most economical option, with a cost of 107€/MWh. 
After that is hydrogen at 121€/MWh, followed by methanol and methane at 143€/MWh and 150€/MWh, respectively.

\begin{figure}[ht!]
    \centering
    \includegraphics[width=0.5\textwidth]{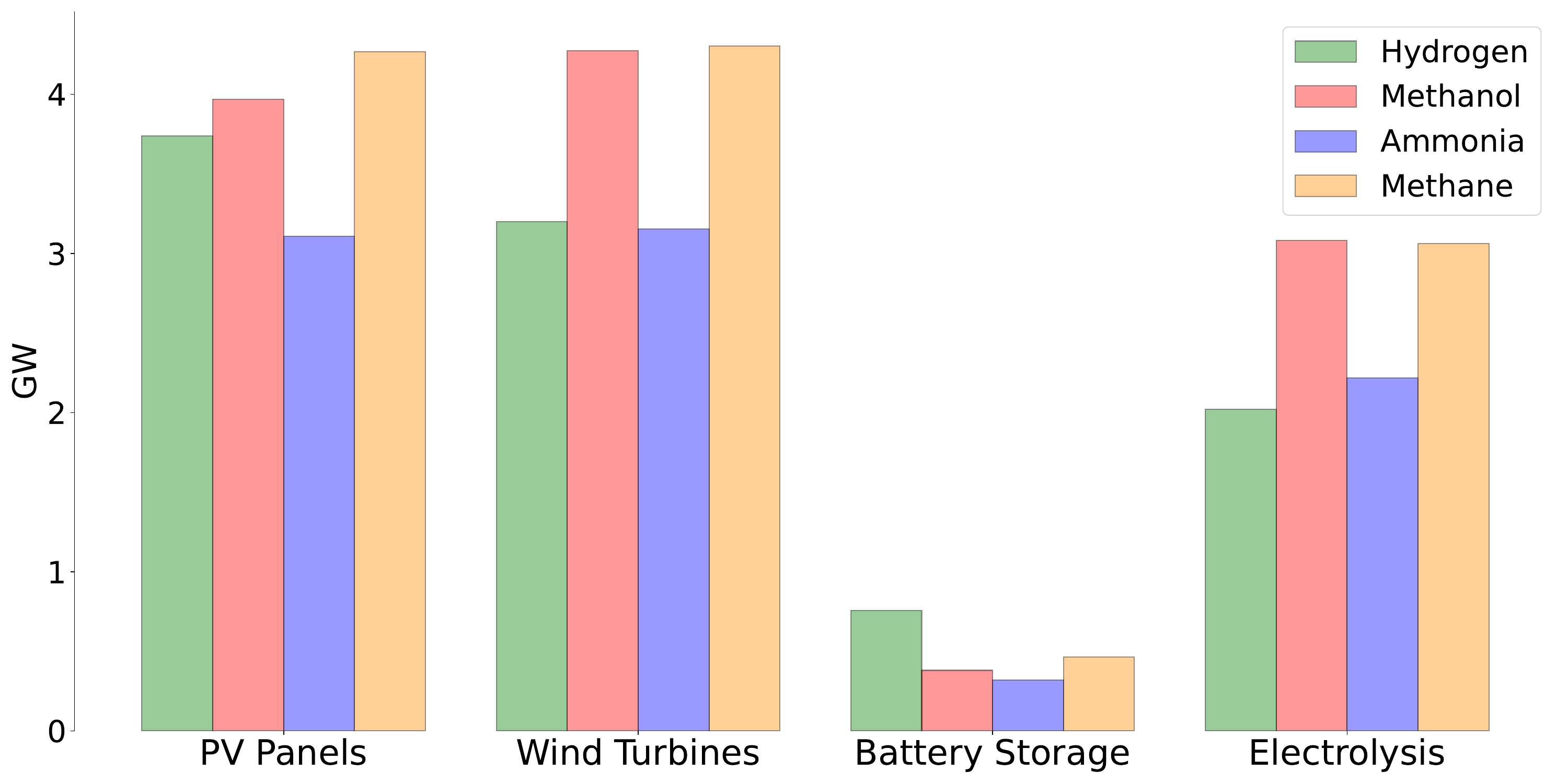}
    \caption{GW-scale capacities from RREHs to meet an annual demand of 10 TWh of e-fuels at the load centre.}
    \label{fig:GW_Install}
\end{figure}

The different installed capacities are depicted in \autoref{fig:GW_Install}. 
The model employing methane as the carrier gas requires the highest installed RE capacity at 8.57 GW for photovoltaic panels and wind turbines. 
In comparaison, the methanol model requires 8.25 GW, the ammonia-based model requires 6.26 GW, and the hydrogen model necessitates 6.94 GW. 
Additionally, the methane and methanol models require 3.06 and 3.08 GW of electrolysers, while the ammonia and hydrogen models require 2.22 GW and 2.0 GW, respectively. 
It is worth mentioning that the power requirements can reflect the overall model efficiency, i.e., the ratio between renewable electricity production and effective e-fuel production (see \autoref{tab:efficiency}).
The overall efficiency is 44\% for methane, 43\% for methanol, 60\% for ammonia, and 55\% for hydrogen.
Most of the energy losses primarily stem from electrolyzers, where efficiency is assumed to be around 75\% \cite{dea2023renewable}, and synthesis units. Specifically, the ammonia synthesis unit achieves an efficiency of 82\% \cite{dea2023renewable}, while that of methanol and methane falls below 80\% \cite{dea2023renewable, gotz2016renewable}. 
Hydrogen liquefaction results in a 30\% \cite{dnv2020study} loss but does not require the use of additional equipment such as DAC (Direct Air Capture) device or an ASU (Air Separation Unit) device.
Concerning battery capacities, the hydrogen model has the highest requirement at 0.76 GW, while the ammonia and methanol models are around 0.3 GW, and the methane model requires 0.46 GW of power. 
It should be noted that the power required for the battery is higher with the hydrogen model. 
This is due to the significant electricity demand needed for the liquefaction process, which must receive a continuous power supply even when electricity production is intermittent.

\begin{table}[!ht]
    \centering
    \begin{tabular}{c|c|c|c|c}
         & NH$_3$ & CH$_3$OH & H$_2$ & CH$_4$ \\
        \hline
        TWh & 0.35 & 0.07 & 0.39 & 0.50 \\
        Km$^3$ & 224.06 & 46.84 & 244.92 & 319.80 \\
    \end{tabular}
    \caption{Storage capacities of gaseous hydrogen at 200 bars and ambient temperature from RREHs for fulfilling an annual 10 TWh e-fuel demand at the load center}
    \label{tab:h2_stock}
\end{table}
Furthermore, it is possible to compare the ability of energy models to accommodate power fluctuations arising from intermittent sources by examining the hydrogen storage requirements (as well as batteries), as presented in \autoref{tab:h2_stock}, with storage expressed in TWh (HHV) and also in m$^3$ considering a density of 40 g/l \cite{dea2023renewable}. 
These compressed gaseous hydrogen storage facilities at 200 bars and room temperature introduce flexibility between electrolyser output and synthesis units.
The demand for compressed hydrogen storage is more substantial for the methane model, requiring over 300,000 m$^3$, while the methanol model necessitates less than 50,000 m$^3$. 
The ammonia and hydrogen models, on the other hand, hover around 230,000 m$^3$. 
Differences in the storage of gaseous hydrogen or in batteries result from the processes attributes as well as commodities requirements (see \autoref{tab:conversion_nodes_tab_tech}). Flexibility parameters are determined by ramp up/down constraints as well as minimum operating levels. Some processes are assumed to operate in a constant mode (Desalination units, DAC, ASU, and Liquefaction units), while others are considered flexible (Electrolysis and Synthesis units).

\begin{table}[!ht]
    \centering
    \begin{tabular}{c|c|c|c|c}
         & NH$_3$ & CH$_3$OH & H$_2$ & CH$_4$ \\
        \hline
        \% & 60 & 43 & 55 & 44 \\
    \end{tabular}
    \caption{Overall efficiency for each RREH which represents the ratio between renewable electricity production and effective e-fuel production.}
    \label{tab:efficiency}
\end{table}

\subsection{Discussion}

This study has extended the one conducted by \citet{berger2021remote} by examining alternative energy carriers. It has demonstrated that the three new RREHs, each with its respective vector (ammonia, hydrogen, and methanol), are all more cost-effective than the methane-based system, with cost savings of 28\%, 20\%, and 5\%, respectively.
It is established that the RREH system utilising ammonia is the most economical system.
Synthesizing ammonia by combining hydrogen with nitrogen, which makes up over 80\% of air composition, outperforms the alternative of pairing hydrogen with carbon dioxide due to its lower concentration in the atmosphere (0.04\%).
This contrast is exemplified in \autoref{fig:COST_EFUEL}, wherein technologies related to the ASU and nitrogen storage account for a mere 2.7\%, whereas DAC and carbon storage in methane and methanol models contribute to around 10\% of the overall system cost.

The hydrogen-based system, while avoiding the need for the aforementioned nitrogen and carbon technologies, necessitates a liquefaction unit resulting in a 30\% energy loss.
The enhanced efficiency inherent in the RREH ammonia system translates into a reduced requirement for installed electric capacity — an imperative aspect given that higher capacity installations correspond to elevated costs, as illustrated in \autoref{fig:COST_EFUEL}.

By comparing the methane and methanol models, the latter stands out due to the absence of a liquefaction (and regasification) unit, especially considering that this unit operates continuously and compels methanation to operate the same way.
This leads to greater storage capacities, as seen with \autoref{tab:h2_stock}.
Furthermore, this requirement for liquefaction facilities translates into an increase in electricity production capacity. The findings shows an higher photovoltaic capacity for the methane production model compared to the methanol production model (see \autoref{fig:GW_Install}).

Using technologies that can adapt to the intermittent characteristics of renewable energy sources, such as flexible liquefaction units, is essential to minimize storage requirements. Storing hydrogen in gaseous form presents a disadvantage due to its inherently low volumetric density.
This characteristic could necessitate substantial storage volumes (depending on the demand from the load centre). Therefore increase the safety risk, particularly considering hydrogen's susceptibility to leaks and high flammability owing to its extensive flammability range and low activation energy \cite{LE20232861}.

\section{Conclusion} \label{sec:conclusion}

In this article, we conducted a comparative study of four e-fuels in the context of RREH (Remote Renewable Energy Hub). 
We examined a specific case related to the production of these e-fuels in Algeria and their deliveries in Belgium. 
Our study was based on the modeling and optimization framework proposed by \citet{berger2021remote}, who had previously studied the production cost of synthetic methane, with an estimated cost of approximately 150 €/MWh (HHV) for the year 2030.

We developed three new RREHs, each using a different e-fuel: ammonia, hydrogen, and methanol. 
This comparative study focused on production costs throughout the supply chain, as well as the technical performance. 
The new models all succeeded in producing their e-fuels at lower costs, respectively 107 €/MWh, 121 €/MWh, and 143 €/MWh (HHV).

The export of ammonia helped reduce costs due to its superior efficiency. 
On one side, by avoiding hydrogen liquefaction, and on the other side, by combining hydrogen with nitrogen rather than carbon dioxide (captured from the air). 
This improved efficiency resulted in a lower installed capacity of RE (PV and wind turbines). To meet an annual demand of 10 TWh of e-fuel (HHV), approximately 6 GW of RE are needed for ammonia and 9 GW for methane (model which requires the most RE).

We also assessed the models' ability to handle fluctuations in intermittent production means by analyzing storage capacities, both in lithium-ion batteries and compressed hydrogen storage. 
The results have shown significant disparities between the models storage needs, with differences of up to six times for compressed hydrogen capacities and approximately three times for batteries.

Moreover, these novel e-fuels, investigated within the framework of RREH, continue to be costlier than their fossil-based counterparts. Nevertheless, considering past energy crises, there is no assurance that this cost advantage will persist over the long term.

Finally, we did not consider structural or parametric uncertainty, nor did we seek to identify near-optimal conditions that could better satisfy other criteria, e.g., environmental, than purely economic ones. Therefore, sensitivity analysis \cite{BORGONOVO2016869} and near-optimal space exploration techniques \cite{DuboisNearOptimality} could be applied in future work to potentially provide additional information to this study.

\section*{Acknowledgements}

This research is supported by the public service of the Belgium federal government (SPF Économie, P.M.E., Classes moyennes et Energie) within the framework of the DRIVER project. The authors express their gratitude to Alexis Costa from the University of Mons for his assistance during the design process of the hubs. Victor Dachet gratefully acknowledges the financial support of the Wallonia-Brussels Federation for his
FRIA grant and the financial support of the Walloon Region for Grant No. 2010235 – ARIAC by DW4AI \hyperlink{https://digitalwallonia4.ai/}{digitalwallonia4.ai}. 

\section*{Glossary}

\begin{itemize}

    \item ASU : Air Separation Unit
    \item CAPEX : Capital Expenditure 
    \item DAC : Direct Air Capture
    \item GBOML : Graph-Based Optimization Modeling Language
    \item HHV : Higher Heating Value
    \item HVDC : High Voltage Direct Current
    \item OPEX : OPerating EXpenditure
    \item PEM : Proton Exchange Membrane
    \item PV : Photovoltaic 
    \item RE : Renewable Energy 
    \item RREH : Remote Renewable Energy Hub
     
\end{itemize}

\section*{Competing interests}
The authors declare no competing interests.

\section*{Declaration of Generative AI and AI-assisted technologies in the writing process}
During the preparation of this work the author(s) used ChatGPT  in order to correct the readiness, grammar and spelling of the writing. After using this tool/service, the authors reviewed and edited the content as needed and take full responsibility for the content of the publication.

\section*{Code and Data Availability}
We would like to emphasize that we open-sourced the code and data used to produce this work \url{https://gitlab.uliege.be/smart_grids/public/gboml/-/tree/master/examples/}

\bibliographystyle{elsarticle-num-names}
\bibliography{bibli}

\onecolumn
\appendix 
% \newpage
\section{Technical parameters used to model conversion nodes} \label{sec:appendix-conversion-nodes-section}

\begin{table}[!ht]
    \centering
    \begin{tabular}{|c|c|c|c|}
    \hline
         & Conversion factor & Minimum & Ramp \\
         &  & level production &  up\&down \\
        \hline
        HVDC Interconnection& 0.9499 & - & - \\
        \cite{ieaETSAP} &  &  & \\
        Electrolysis & 50.6 GWh$_{e_l}$/kt$_{H_2}$ & 0.05 & 1.0 /h \\
        \cite{gotz2016renewable} & 9.0 kt$_{H_2O}$/kt$_{H_2}$ & & \\
        Methanation & 0.5 kt$_{H_2}$/kt$_{CH_4}$ & 0.4 & 1.0 /h \\
        \cite{gotz2016renewable} \cite{RONSCH2016276} & 2.75 kt$_{CO_2}$/kt$_{CH_4}$ &  &  \\
         & 2.25 kt$_{H_20}$/kt$_{CH_4}$ & & \\
        Haber-Bosch & 0.32 GWh$_{e_l}$/kt$_{NH_3}$  & 0.2 & 1.0 /h\\
        \cite{dea2023renewable} & 0.18 kt$_{H_2}$/kt$_{NH_3}$ &  & \\
         & 0.84 kt$_{N_2}$/kt$_{NH_3}$ & & \\
        e-CH$_3$OH  & 0.1 GWh$_{e_l}$/kt$_{CH_3OH}$ & 0.1 &  1.0 /h\\
        \cite{dea2023renewable} & 0.209 kt$_{H_2}$/kt$_{CH_3OH}$   &  & \\
         & 1.37 kt$_{CO_2}$/kt$_{CH_3OH}$ & & \\
         & 0.93 kt$_{H_2O}$/kt$_{CH_3OH}$ & & \\
        Desalination & 0.004 GWh$_{e_l}$/kt$_{H_2O}$ & 1.0 & 0.0 /h \\
        \cite{irenawaterdessal} &  &  &  \\
        Direct Air Capture  & 0.1091 GWh$_{e_l}$/kt$_{CO_2}$ & 1.0 & 0.0 /h\\
        \cite{KEITH20181573} & 0.0438 H$_{2}$/kt$_{CO_2}$  & & \\
         & 5.0 H$_{2}$O/kt$_{CO_2}$ & & \\
        Air Separation Unit & 0.1081 GWh$_{e_l}$/kt$_{N_2}$ & 1.0 & 0.0 /h\\
        \cite{morgan2013technoeconomic} &  &  & \\
        CH$_4$ Liquefaction & 10.616 GWh$_{e_l}$/kt$_{LCH_4}$ & 1.0 & 0.0 /h\\
        \cite{POSPISIL20191} &  &  & \\
        H$_2$ Liquefaction  & 12 GWh$_{e_l}$/kt$_{H_2}$ & 1.0 & 0.0 /h \\
        \cite{connellyH2} &  &  & \\
        LCH$_4$ Carriers  & 0.994 & - & -\\
        \cite{rogersLNG} &  &  & \\
        LH$_2$ Carriers  & 0.945 & - & -\\
        \cite{dea2023energytransport} &  &  & \\
        LNH$_3$ Carriers  & 0.994 & - & -\\
        \cite{dea2023energytransport} &  &  & \\
        CH$_3$OH Carriers & 0.993 & - & -\\
        \cite{dea2023energytransport} &  &  & \\
        LCH$_4$ Regasification  & 0.98 & - & - \\
        \cite{POSPISIL20191} &  &  & \\
        LNH$_3$ Regasification & 0.98 & - & -\\
        Assumed &  &  & \\
        LH$_2$ Regasification  & 1.0 & - & -\\
        \cite{dnv2020study}&  &  & \\
    \hline
    \end{tabular}
    \caption{Technical parameters used for modeling conversion nodes (2030 estimate)}
    \label{tab:conversion_nodes_tab_tech}
\end{table}

\newpage
\section{Technical parameters used to model storage nodes} \label{sec:appendix-storage-nodes_section}

\begin{table}[!ht]
    \centering
    %\begin{adjustbox}{angle=90}
    \begin{tabular}{|c|c|c|c|c|c|c|}
    \hline
        & Conversion Factor &  Self-discharge & Charge & Discharge & Discharge-to & Minimum\\
        &  & Rate & Efficiency & Efficiency & -charge ratio & inventory level\\
        \hline
        Battery Storage & - & 0.00004 & 0.959 &  0.959 & 1.0 & 0.0\\
        \cite{dea2023energystorage} &  &  &  &  &  & \\
        H$_2$ Storage (200 bars) & 1.3 GWh$_{e_l}$/kt$_{H_2}$ & 1.0 & 1.0 & 1.0 & 1.0 & 0.05\\
        \cite{dea2023energystorage} &  &  &  &  &  & \\
        Liquefied $CO_2$ Storage & 0.105 GWh$_{e_l}$/kt$_{CO_2}$ & 1.0 & 1.0 & 1.0 & 1.0 & 0.0\\
        \cite{mistubishiCO2} &  &  &  &  &  & \\
        Liquefied CH$_4$ Storage & - & 1.0 & 1.0 & 1.0 & 1.0 & 0.0\\
        Assumed &  &  &  &  &  & \\
        Liquefied NH$_3$ Storage & - & 0.00003 & 1.0 & 1.0 & 1.0 & 0.0\\
        \cite{dnv2020study}&  &  &  &  &  & \\
        Liquefied H$_2$ Storage & - & 0.00008 & 1.0 & 1.0 & 1.0 & 0.0\\
        \cite{iea2019}&  &  &  &  &  & \\
        CH$_3$OH Storage & - & 1.0 & 1.0 & 1.0 & 1.0 & 0.0\\
        \cite{dea2023energystorage}&  &  &  &  &  & \\
        $H_2O$ Storage & 0.00036 GWh$_{e_l}$/kt$_{H_20}$ & 1.0 & 1.0 & 1.0 & 1.0 & 0.0\\
        \cite{CALDERA2016207}&  &  &  &  &  & \\
        $N_2$ Storage & 0.1081 GWh$_{e_l}$/kt$_{N_2}$ & 1.0 & 1.0 & 1.0 & 1.0 & 0.0 \\
        Assumed &  &  &  &  &  & \\
        
    \hline
    \end{tabular}
    %\end{adjustbox}
    \caption{Technical parameters used for modeling storage nodes (2030 estimate)}
    \label{tab:Storage_nodes_tab_tech}
\end{table}

\newpage
\section{Economic parameters used to model conversion nodes } \label{sec:appendix-conversion-nodes_sec_eco}

\begin{table}[!ht]
    \centering
    \begin{tabular}{|c|c|c|c|c|}
    \hline
         & CAPEX & FOM & VOM & Lifetime \\
        \hline
        Solar Photovoltaic Panels & 380.0 M€/$GW_{el}$ & 7.25 M€/$GW_{el}-yr$ & 0.0 M€/$GWh$ & 25.0 yr\\
        \cite{dea2023energyproduction} & & & & \\
        Wind Turbines & 1040.0 M€/$GW_{el}$ & 12.6 M€/$GW_{el}-yr$ & 0.00135 M€/$GWh$ & 30.0 yr\\
        \cite{dea2023energyproduction} & & & & \\
        HVDC Interconnection & 480.0 M€/$GW_{el}$ & 7.1 M€/$GW_{el}-yr$ & 0.0 M€/$GWh$ & 40.0 yr\\
        \cite{eiatransmission}& & & & \\
        Electrolysis & 600.0 M€/$GW_{el}$ & 30 M€/$GW_{el}-yr$ & 0.0 M€/$GWh$ & 15.0 yr\\
        \cite{dea2023renewable}& & & & \\
        Methanation & 735.0 M€/$GW_{CH_4}$ & 29.4 M€/$GW_{CH_4}-yr$ & 0.0 M€/$GWh_{CH_4}$ & 20.0 yr\\
        \cite{iea2019}& & & & \\
        Haber-Bosch & 6825.0 M€/$kt_{NH_3}/h$ & 204.75 M€/$kt_{NH_3}/h-yr$ & 0.000105 M€/$kt_{NH3}$ & 30.0 yr\\
        \cite{dea2023renewable}& & & & \\
        e-CH$_3$OH & 57252.80 M€/$kt_{CH_3OH}/h$ & 158.31 M€/$kt_{CH_3OH}/h-yr$ & 0.0 M€/$kt_{CH3OH}$ & 30.0 yr\\
        \cite{dea2023renewable}& & & & \\
        Desalination & 28.08 M€/$kt_{H_2O}/h$ & 0.0 M€/$kt_{H_20}/h-yr$ & 0.000315 M€/$kt_{H2O}$ & 20.0 yr\\
        \cite{cmimarseille}& & & & \\
        Direct Air Capture & 4801.4 M€/$kt_{CO_2}/h$ & 0.0 M€/$kt_{C0_2}/h-yr$ & 0.0207 M€/$kt_{C02}$ & 30.0 yr\\
        \cite{KEITH20181573}& & & & \\
        Air Separation Unit & 850.0 M€/$kt_{N_2}/h$ & 50.0 M€/$kt_{N_2}/h-yr$ & 0.0 M€/$kt_{N2}$ & 30.0 yr\\
        \cite{morgan2013technoeconomic}& & & & \\
        CH$_4$ Liquefaction & 5913.0 M€/$kt_{LCH_4}/h$ & 147.825 M€/$kt_{LCH_4}/h-yr$ & 0.0 M€/$kt_{LCH4}$ & 30.0 yr\\
        \cite{songhurstlng}& & & & \\
        H$_2$ Liquefaction & 45000.0 M€/$kt_{LH_2}/h$ & 1125.825 M€/$kt_{LH_2}/h-yr$ & 0.0 M€/$kt_{LH2}$ & 40.0 yr\\
        \cite{dnv2020study}& & & & \\
        LCH$_4$ Carriers & 2.537 M€/$kt_{LCH_4}/h$ & 0.12685  M€/$kt_{LCH_4}/h-yr$ & 0.0 M€/$kt_{LCH4}$ & 30.0 yr\\
        \cite{ERIAship}& & & & \\
        LH$_2$ Carriers & 14.0 M€/$kt_{LH_2}/h$ & 0.07  M€/$kt_{LH_2}/h-yr$ & 0.0 M€/$kt_{LH2}$ & 30.0 yr\\
        \cite{dea2023energytransport}& & & & \\
        LNH$_3$ Carriers & 1.75 M€/$kt_{LNH_3}/h$ & 0.09  M€/$kt_{LNH_3}/h-yr$ & 0.0 M€/$kt_{LNH3}$ & 30.0 yr\\
        \cite{dea2023energytransport}& & & & \\
        CH$_3$OH Carriers & 0.69 M€/$kt_{CH_3OH}/h$ & 0.04 M€/$kt_{CH_3OH}/h-yr$ & 0.0 M€/$kt_{CH3OH}$ & 30.0 yr\\
        \cite{dea2023energytransport}& & & & \\
        LCH$_4$ Regasification & 1248.3 M€/$kt_{LCH_4}/h$ & 29.97 M€/$kt_{LCH_4}/h-yr$ & 0.0 M€/$kt_{LCH4}$ & 30.0 yr\\
        \cite{Dongsha2017}& & & & \\
        LNH$_3$ Regasification & 1248.3 M€/$kt_{LNH_3}/h$ & 29.97 M€/$kt_{LNH_3}/h-yr$ & 0.0 M€/$kt_{LNH3}$ & 30.0 yr\\
        Assumed & & & & \\
        LH$_2$ Regasification & 9100.0 M€/$kt_{LH_2}/h$ & 27.8 M€/$kt_{LH_2}/h-yr$ & 0.0 M€/$kt_{LH2}$ & 30.0 yr\\
        \cite{dnv2020study}& & & & \\
    \hline
    \end{tabular}
    \caption{Economical parameters used for modeling conversion nodes (2030 estimate)}
    \label{tab:conversion_nodes_tab_eco}
\end{table}

\newpage
\section{Economic parameters used to model storage nodes (flow component)} \label{sec:appendix-storage-nodes_sec_eco}

\begin{table}[!ht]
    \centering
    %\begin{adjustbox}{angle=90}
    \begin{tabular}{|c|c|c|c|c|}
    \hline
         & CAPEX &  FOM & VOM & Lifetime\\
        \hline
        Battery Storage & 160.0 M€/GW & 0.5 M€/GW-yr & 0.0 M€/GWh & 10.0 yr \\
        \cite{dea2023energystorage}& & & & \\
        Liquefied $CO_2$ Storage & 48.6 M€/(kt/h) & 2.43 M€/(kt/h)-yr & 0.0 M€/kt & 30 yr\\
        \cite{mistubishiCO2}& & & & \\
        Liquefied NH$_3$ Storage & 0.10 M€/(kt/h) & 0.001 M€/(kt/h)-yr & 0.0 M€/kt & 30 yr\\
        \cite{dea2023renewable}& & & & \\
        CH$_3$OH Storage & 0.0625 M€/(kt/h) & 0.0 M€/(kt/h)-yr & 0.0 M€/kt & 30 yr\\
        \cite{dea2023energystorage}& & & & \\
        $H_2O$ Storage & 1.55923 M€/(kt/h) & 0.0312 M€/(kt/h)-yr & 0.0 M€/kt & 30 yr\\
        \cite{CALDERA2016207}& & & & \\
        
    \hline
    \end{tabular}
    %\end{adjustbox}
    \caption{Economic parameters used to model storage nodes (flow component)}
    \label{tab:Storage_nodes_tab_eco_flow}
\end{table}

\section{Economic parameters used to model storage nodes (stock component)} \label{sec:appendix-storage_nodes_sec_stock}

\begin{table}[!ht]
    \centering
    %\begin{adjustbox}{angle=90}
    \begin{tabular}{|c|c|c|c|c|}
    \hline
         & CAPEX &  FOM & VOM & Lifetime\\
        \hline
        Battery Storage & 142.0 M€/GWh & 0.0 M€/GWh-yr & 0.0018 M€/GWh & 10.0 yr \\
        \cite{dea2023energystorage}& & & & \\
        H$_2$ Storage (200 bars) & 45.0 M€/kt & 2.25 M€/kt-yr & 0.0 M€/kt & 30 yr\\
        \cite{dea2023energystorage}& & & & \\
        Liquefied $CO_2$ Storage & 1.35 M€/kt & 0.0675 M€/kt-yr & 0.0 M€/kt & 30 yr\\
        \cite{mistubishiCO2}& & & & \\
        Liquefied CH$_4$ Storage & 2.641 M€/kt & 0.05282 M€/kt-yr & 0.0 M€/kt & 30 yr\\
        \cite{iguLNG}& & & & \\
        Liquefied NH$_3$ Storage & 0.867 M€/kt & 0.01735 M€/kt-yr & 0.0 M€/kt & 30 yr\\
        \cite{morgan2013technoeconomic}& & & & \\
        Liquefied H$_2$ Storage & 25.0 M€/kt & 0.5 M€/kt-yr & 0.0 M€/kt & 30 yr\\
        \cite{dnv2020study}& & & & \\
        CH$_3$OH Storage & 2.778 M€/kt & 0.0 M€/kt-yr & 0.0 M€/kt & 30 yr\\
        \cite{dea2023energystorage}& & & & \\
        $H_2O$ Storage & 0.065 M€/kt & 0.0013 M€/kt-yr & 0.0 M€/kt & 30 yr\\
        \cite{CALDERA2016207}& & & & \\
        $N_2$ Storage & 45.0 M€/kt & 2.25 M€/kt-yr & 0.0 M€/kt & 30 yr\\
        Assumed & & & & \\
        
    \hline
    \end{tabular}
    %\end{adjustbox}
    \caption{Economic parameters used to model storage nodes (stock component)}
    \label{tab:Storage_nodes_stock_tab_eco_stock}
\end{table}

\end{document}